\documentclass[10pt]{article}
\usepackage{a4wide}
\usepackage{amstext,amsfonts,amsmath,theorem}
\usepackage{graphicx}

\begin{document}

\thispagestyle{empty}

\begin{center}

\begin{Large}
{\bf Extensions, expansions, Lie algebra cohomology and enlarged
superspaces\footnote{Invited talk delivered at the {\it EU RTN
Workshop}, Copenhagen, Sep.~15-19 and at the {\it Argonne Workshop
on Branes and Generalized Dynamics}, Oct.~20-24, 2003.}}
\end{Large}
\vskip 1cm

\begin{large}
J.~A.~de~Azc\'arraga$^{a}$, J.~M. Izquierdo$^{b}$,
M.~Pic\'on$^{a}$ and O.~Varela$^{a}$
\end{large}
\vspace*{0.6cm}

\begin{it}
$^a$ Departamento de F\'{\i}sica Te\'orica, Facultad de
F\'{\i}sica,
Universidad de Valencia\\
and IFIC, Centro Mixto Universidad de Valencia--CSIC,
\\
E--46100 Burjassot (Valencia), Spain
\\
$^b$ Departamento de F\'{\i}sica Te\'orica, Universidad de
Valladolid
\\
E--47011 Valladolid, Spain
\\[0.4cm]
\end{it}
\end{center}

\begin{abstract}
After briefly reviewing the methods that allow us to derive
consistently new Lie (super)algebras from given ones, we consider
enlarged superspaces and superalgebras, their relevance and some
possible applications.
\end{abstract}

\section{Introduction}
\label{int}

Superstring theories, and their low-energy supergravity limits,
have made apparent that the original supersymmetry algebra has to
be enlarged beyond the restrictions imposed by the
Haag-$\L$opusza\'nski-Sohnius theorem \cite{HLS}. This was so, in
particular, for the following reasons:

$\bullet$ $D=11$ supergravity \cite{CJS} may be formulated in a
way \cite{DFR,CFGPvN} which suggests that its possible underlying
(gauge) group is related to $OSp(1|32)$ (see also
\cite{Zanelli,HN} and references therein).

$\bullet$ In situations where the topology is non-trivial, the
quasi-invariance under supersymmetry of the Wess-Zumino (WZ) terms
of the brane actions results in algebras realized by the conserved
supercharges that include additional (topological) charges and
that are extensions of the original supersymmetry algebra
\cite{AGIT}.

$\bullet$ The existence of solitonic brane solutions of the
different supergravities that preserve a fraction of the
supersymmetry may be explained from an algebraic point of view by
considering more general forms of the algebra (see \cite{To,Duff}
and references therein), and described by the preon hypothesis
\cite{BAIL} (see also \cite{BAIPV}).

The lesson to be learnt from these facts is that wherever there is
a consistent modification of a given symmetry algebra, it will
probably show up in an application. This spirit, in fact, inspired
the old search for mixed unitary and kinematical symmetries that
was halted by the well known no-go theorems (see \cite{Dyson} for
a history of the subject), theorems that were finally bypassed by
the realization that fermionic symmetries should be included, and
hence by supersymmetry. In fact, if one grants that fermionic
spinors exist as the only
 primary entities, already ordinary supersymmetry is seen
to be a natural outcome: it is the result of a central extension
of the odd abelian spinor translation group by the group of
spacetime translations \cite{AA85}. Thus, it makes sense to search
for supersymmetry algebras beyond the standard superPoincar\'e
algebra (see \cite{vHvP,DFR,AGIT,SB,CAIPB00,AI01} and references
therein).

With this point of view, we shall first review the known methods
for obtaining new algebras from given ones, {\it
i.e.}~contractions, deformations and extensions of Lie and super
Lie algebras, plus a new one (that {\it includes} contractions)
which we have called in \cite{AIPV} Lie (super)algebra expansions.
Next we shall concentrate on extensions and expansions, and look
for physical applications in both cases.

\section{Four ways to relate and derive Lie (super)algebras}
\label{fourw}

{\it (a)  Contractions of Lie (super)algebras}

In their simplest \.In\"on\"u-Wigner (IW) form \cite{IW53}, the
contraction of $\mathcal{G}$ with respect to a subalgebra
$\mathcal{L}_0 \subset {\cal G}$ is performed by rescaling the
generators of the coset $\mathcal{G}/\mathcal{L}_0$, and then by
taking a singular limit for the rescaling parameter.

This procedure can be extended to generalized IW contractions in
the sense of Weimar-Woods (W-W) \cite{Wei:00}. These are defined
when $\mathcal{G}$ can be split in a sum of vector subspaces
\begin{equation}
      \mathcal{G}= V_0\oplus V_1\oplus \dots\oplus V_n = \bigoplus_{s=0}^n V_s ,
\label{gc1}
\end{equation}
($V_0$ being the vector space of the subalgebra $\mathcal{L}_0$),
such that the following conditions are satisfied:
\begin{equation}
\hspace{-1.5cm}    c^{k_s}_{i_pj_q}= 0 \;\; \text{if} \ s>p+q
\qquad \text{i.e.}\qquad  [V_p,V_q]\subset \bigoplus_s V_s,\ s\leq
p+q \ , \label{gc2}
\end{equation}
where $i_p$ labels the generators of $\mathcal{G}$ in $V_p$, and
$c^k_{ij}$ are structure constants of $\mathcal{G}$. Then the W-W
\cite{Wei:00} contracted algebra is obtained by rescaling the
group parameters as $g^{i_p}\mapsto \lambda^p g^{i_p}$,
$p=0,\dots,n$ and then by taking a singular limit for $\lambda$.
The contracted Lie algebra obtained this way, $\mathcal{G}_c$, has
the same dimension as $\mathcal{G}$. The case $n=1$ corresponds to
the simple IW contraction.

Well known examples of contractions that appear in physics include
the Galilei algebra as an IW contraction of the Poincar\'e
algebra, the Poincar\'e algebra as a contraction of the de Sitter
algebras, or the characterization of the M-theory superalgebra
\cite{To} (ignoring the Lorentz part) as a contraction of
$osp(1|32)$.

\medskip

\noindent {\it (b) Deformations}

Lie algebra deformations \cite{DEF} can be regarded, from the
physical point of view, as a process inverse to contractions.
Mathematically, a deformation $\mathcal{G}_d$ of a Lie algebra
$\cal{G}$ is a Lie algebra close, but not isomorphic, to
$\mathcal{G}$. As in the case of contractions, $\mathcal{G}_d$ has
the same dimension as $\mathcal{G}$.

Deformations are performed by modifying the r.h.s.~of the original
commutators by adding new terms that depend on a parameter $t$ in
the form
\begin{equation}
\hspace{-1cm}  [X,Y]_t =[X,Y]_0+\sum^\infty_{i=1} \omega_i(X,Y)
t^i \ , \quad X,Y \in {\cal G}\;,\quad \omega_i(X,Y) \in {\cal G}
\; . \label{def1}
\end{equation}
Checking the Jacobi identities up to $O(t^2)$, it is seen that the
expression satisfied by $\omega_1$ characterizes it as a
two-cocycle so that the second Lie algebra cohomology group
$H^2(\mathcal{G}, \mathcal{G})$ of $\mathcal{G}$ with coefficients
in the Lie algebra $\mathcal{G}$ itself is the group of
infinitesimal deformations of $\mathcal{G}$. Thus
$H^2(\mathcal{G}, \mathcal{G})=0$ is a sufficient condition for
rigidity \cite{DEF}. In this case, $\mathcal{G}$ is {\it rigid} or
{\it stable} under infinitesimal deformations; any attempt to
deform it yields an isomorphic algebra. The problem of finite
deformations depends on the integrability condition of the
infinitesimal deformation; the obstruction is governed by the
third cohomology group $H^3(\mathcal{G}, \mathcal{G})$ that needs
being trivial.

As is known, the Poincar\'e algebra may be seen as  a deformation
of the Galilei algebra, a fact that may be viewed as a group
theoretical prediction of relativity; $so(4,1)$ and $so(3,2)$ are
stabilizations of the Poincar\'e algebra; $osp(1|4)$ is a
deformation of the $N=1$, $D=4$ superPoincar\'e algebra (for
deformations of Lie superalgebras see \cite{B86}). Nontrivial
central extensions (see (c) below) of Lie algebras may also be
considered as deformations or partial stabilizations of trivial
ones.

\medskip

\noindent{\it (c) Extensions}

In contrast with the procedures (a), (b), the initial data of the
extension problem include {\it two} algebras ${\cal G}$ and ${\cal
A}$. A Lie algebra $\tilde{\mathcal{G}}$ is an extension  of the
Lie algebra $\mathcal{G}$ by the Lie algebra $\mathcal{A}$ if
$\mathcal{A}$ is an ideal of $\tilde{\mathcal{G}}$ and
$\tilde{\mathcal{G}}/\mathcal{A}=\mathcal{G}$. As a result,
$\text{dim}\, \tilde{\mathcal{G}}= \text{dim}\, \mathcal{G} +
\text{dim}\, \mathcal{A}$, so that this process is also `dimension
preserving'.

Given $\mathcal{G}$ and $\mathcal{A}$, in order to obtain an
extension $\tilde{\mathcal{G}}$ of $\mathcal{G}$ by $\mathcal{A}$
it is necessary to specify first an action $\rho$ of $\mathcal{G}$
on $\mathcal{A}$ {\it i.e.}, a Lie algebra homomorphism $\rho: \,
\mathcal{G} \longrightarrow \text{End}\, \mathcal{A}$. The
different possible extensions $\tilde{\mathcal{G}}$ for
$(\mathcal{G},\mathcal{A} ,\rho)$ and the possible obstructions to
the extension process are, once again, governed by cohomology
\cite{EXT}. To be more explicit, let $\mathcal{A}$ be abelian. The
extensions are governed by $H^2_\rho(\mathcal{G},\mathcal{A})$.
Some special cases are: 1) trivial action $\rho=0$,
$H^2_0(\mathcal{G},\mathcal{A})\neq 0$. These are central
extensions, in which $\mathcal{A}$ belongs to the centre of
$\tilde{\mathcal{G}}$; they are determined by non-trivial ${\cal
A}$-valued two-cocycles on ${\cal G}$, and non-equivalent
extensions correspond to non-equivalent cocycles; 2) non-trivial
action $\rho\neq 0$, $H^2_\rho(\mathcal{G},\mathcal{A})= 0$
(semidirect extension of $\mathcal{G}$ by $\mathcal{A}$); and 3)
$\rho=0$, $H^2(\mathcal{G},\mathcal{A})=0$ (direct sum of
$\mathcal{G}$ and $\mathcal{A}$, $\tilde{\mathcal{G}} =
\mathcal{G} \oplus \mathcal{A}$, or trivial extension).

Well-known examples of extensions in physics are the centrally
extended Galilei algebra, which is relevant in quantum mechanics;
the two-dimensional extended Poincar\'e algebra that allows
\cite{CJ92} for a gauge theoretical derivation of the
Callan-Giddings-Harvey-Strominger model, or the M-theory
superalgebra that, without the Lorentz automorphisms part, is the
maximal central extension of the abelian $D=11$ supertranslations
algebra (see Sec.~\ref{FMthe} and \cite{vHvP,To,CAIPB00}).

\medskip

\noindent {\it (d) Expansions}

Under a different name, Lie algebra expansions were first used in
\cite{HS}, and then the method was studied in general in
\cite{AIPV}. The idea is to consider the Maurer-Cartan (MC)
equations of the starting Lie algebra $\mathcal{G}$ in terms of
the invariant forms on the group manifold, and then perform a
rescaling of some of the group parameters $g^i$,
$i=1,\dots,\text{dim}\, \mathcal{G}$, by a parameter $\lambda$.
Then, one expands the invariant one-forms $\omega^i$ in $\lambda$.
Inserting these expansions (polynomials in $\lambda$) in the
original MC equations for $\mathcal{G}$,
\begin{equation}
             d\omega^i=-\frac{1}{2} {c^i}_{jk}
\omega^j\wedge \omega^k \ , \label{MC}
\end{equation}
one obtains a set of equations that have to be satisfied, one for
each power of $\lambda$. The problem to be addressed then is how
to cut the series expansions of the $\omega^i$'s in such a way
that the resulting set of equations remains consistent {\it i.e.},
closed under $d$, so that it defines the MC equations of a new
algebra, the {\it expanded} Lie algebra. We do not enumerate all
the possibilities here \cite{AIPV}. We shall just mention that,
under the W-W conditions \cite{Wei:00} for generalized
contractions, Eq.~(\ref{gc2}), and with the corresponding
rescaling, the $\{\omega^i\}$ MC forms are divided into $n+1$ sets
$\{\omega^{i_p}\}$, $p=0,1,\dots n$, and the forms $\omega^{i_p}$
corresponding to each subspace in (\ref{gc1}) have the expansion
\begin{equation}
\hspace{-1.5cm} \omega^{i_p} = \sum^\infty_{s=p}
\omega^{i_p,s}\lambda^s \;, \qquad {\mathrm i.e.} \qquad
\omega^{i_p}(\lambda) = \lambda^p \omega^{i_p , p} + \lambda^{p+1}
\omega^{i_p , p+1} + \ldots \label{expWW}
\end{equation}
(see \cite{AIPV}). If one demands that the maximum power in the
expansion of the forms $\{\omega^{i_p}\}$ in the $p$--th subspace
is $N_p\geq p$, then consistency requires that
\begin{equation}
\hspace{-1.5cm} N_{q+1} = N_q \qquad \text{or} \qquad N_{q+1} =
N_q +1 \qquad (q=0,1,\ldots, n-1)\ . \label{conWW}
\end{equation}
The new Lie algebras, generated by the MC forms
\begin{equation}
\hspace{-1.5cm} \{ \omega^{i_{ 0}, 0}, \omega^{i_{ 0}, 1},
\stackrel{N_0 +1}{\ldots}, \omega^{i_{ 0}, N_{ 0}};\, \,
\omega^{i_{ 1}, 1}, \stackrel{N_1}{\ldots}, \omega^{i_{ 1}, N_{
1}}; \,\, \ldots; \,\, \omega^{i_{n}, n},
\stackrel{N_n-n+1}{\ldots}, \omega^{i_{n}, N_{n}} \} \; ,
\end{equation}
are labelled $\mathcal{G}(N_0,N_1,\dots,N_n)$ and define {\it
expansions} of the original Lie algebra $\mathcal{G}$. The case
$N_p=p \,$, $\mathcal{G}(0,1,\dots,n)$, coincides with the
generalized W-W contraction and has the same dimension as the
original ${\cal G}$; thus, the W-W contraction is a particular
expansion. In all other cases the expanded algebra
$\mathcal{G}(N_0,N_1,\dots,N_n)$ is larger than $\mathcal{G}$
[$\textrm{dim} \, \mathcal{G}(N_0, \ldots, N_n) = \sum_{p=0}^{n}
(N_p -p+1) \, \textrm{dim} \, V_p$], so that the expansion process
is not `dimension preserving' (hence its name).

Other interesting cases are those of Lie superalgebras with
splittings satisfying the W-W conditions {\it e.g.}, of the form
$\mathcal{G}=V_0\oplus V_1$ or $\mathcal{G}=V_0 \oplus V_1 \oplus
V_2$ and such that $V_0$ or $V_0\oplus V_2$ contain all the
bosonic generators and $V_1$ contains the fermionic ones. Then,
the expansions of the one-forms in the (dual) subspaces $V_1^*$
($V_0^*$ and $V_2^*$) of ${\cal G}^*$ only contain odd (even)
powers of $\lambda$. The consistency conditions for the existence
of the $\mathcal{G}(N_0,N_1)$ and $\mathcal{G}(N_0,N_1,N_2)$
expanded superalgebras require that
\begin{equation}
    N_0=N_1-1\ , \ N_0=N_1+1 \quad ,
\label{consup1}
\end{equation}
and
\begin{equation} \hspace{-1.5cm}
  N_0=N_1+1=N_2\ , \ \;
N_0=N_1-1=N_2\ , \ \; N_0= N_1-1=N_2- 2 \quad , \label{consup2}
\end{equation}
respectively.

\section{Super--$p$--branes and extended superspaces with
additional fermionic generators} \label{Nfermionic}

As mentioned, the standard supersymmetry algebra $\{ Q_\alpha,
Q_\beta\} =(C\Gamma^\mu)_{\alpha\beta} P_\mu$,
$[Q_\alpha,P_\mu]=0$, may be viewed  \cite{AA85} as a central
extension of the odd abelian algebra $\{ Q_\alpha, Q_\beta\} =0$
by the spacetime translations. Other `central' (ignoring the
Lorentz part) extensions, with additional bosonic generators, are
realized in brane theory and have a topological origin, as shown
in \cite{AGIT}. Thus, one may ask whether modifying the $[Q,P]$
commutator by adding new fermionic generators also gives
physically relevant supersymmetry algebras.

The first example was the Green algebra \cite{Green}, which
contains an additional fermionic generator, $Z_\alpha$, that
extends centrally the graded translations algebra (superPoincar\'e
without the Lorentz part) provided that the gamma matrices obey an
identity that is satisfied only for the number of spacetime
dimensions for which superstrings exist. Further examples were
given in  \cite{BeSe,Malg}, which gave the form of the spacetime
superalgebras underlying the Lie algebra cohomology
characterization \cite{AT89} of the WZ terms of the scalar
$p$-branes \cite{AETW}. This led naturally to the consideration of
enlarged superspaces that may be seen to have a supergroup
extension structure \cite{CAIPB00}. Using them, it is possible to
construct the super--$p$--brane actions in such a way that the WZ
terms become strictly invariant: then, the Chevalley-Eilenberg
(CE) Lie algebra cohomology $(2p+2)$-cocycles that define the WZ
terms of the scalar $p$--branes \cite{AT89} are trivialized
\cite{BeSe,CAIPB00} (for further work along this line see
\cite{HH,AG03}). These algebras are not central extensions of the
starting centrally extended algebra,
\begin{equation}
\hspace{-2.5cm} \{ Q_\alpha, Q_\beta\}
=(C\Gamma^\mu)_{\alpha\beta} P_\mu
+(C\Gamma^{\mu_1\dots\mu_p})_{\alpha\beta}Z_{\mu_1\dots\mu_p} \, ,
\;\; [Q_\alpha,P_\mu]=0=[Q_\alpha, Z_{\mu_1\ldots\mu_p}] \;\; ,
\label{pbralg}
\end{equation}
although they can be obtained by a step by step process by
extending centrally the previous one. In the fist step, one
extends centrally (\ref{pbralg}) by adding the  new fermionic
generators $Z_{\mu_1\dots\mu_{p-1}\alpha_1}$ (the case for the
Green algebra corresponds to $p=1$). The resulting algebra can be
extended again centrally by  bosonic generators of the form
$Z_{\mu_1\dots\mu_{p-2}\alpha_1\alpha_2}$; this yields an algebra
that is not a central extension of the original one,
Eq.~(\ref{pbralg}). The procedure continues
\cite{BeSe,Malg,CAIPB00} by adding centrally more generators of
the type $Z_{\mu_1\dots\mu_{p-k}\alpha_1\dots\alpha_k}$, and it
ends when one reaches a set of generators where all spacetime
indices have been replaced by spinorial ones. Interestingly
enough, the existence of these extensions depends on the same
gamma matrix identities valid for the $(D,p)$ values that allow
for the existence of the given super--$p$--brane.

Although the new, extended superspaces (generically denoted
$\widetilde \Sigma$) trivialize the WZ terms of the $p$--brane
actions, their relevance, beyond the topologically non-trivial
case, is marginal here since the new superspace group variables
corresponding to the new superalgebra generators appear in the
action (in the WZ term) through a total derivative, and therefore
they do not modify the Euler-Lagrange equations (it will be
different for the D$p$-branes case below). Let us see this more
explicitly. The Lagrangian density of a scalar $p$--brane is of
the form $\mathcal{L}=\mathcal{L}_0+\mathcal{L}_{WZ}$, where
$\mathcal{L}_0$ is the kinetic part and $\mathcal{L}_{WZ}$ is the
WZ term, given by $\mathcal{L}_{WZ}d^{p+1}\xi=\phi^* b$, where
$\phi$ is the mapping that locates the $p$--brane in rigid
superspace, and $b$ is a $(p+1)$-form such that
\begin{equation}
       h=db\propto \Pi^\alpha\wedge
(C\Gamma_{\mu_1\dots\mu_p})_{\alpha \beta}\Pi^\beta \wedge
\Pi^{\mu_1}\wedge \dots \wedge \Pi^{\mu_p}\ , \label{bpbrane}
\end{equation}
$\Pi^\alpha$ and $\Pi^\mu$ being, in the standard flat superspace
$\Sigma$ parametrized by $(x^\mu, \theta^\alpha)$, the invariant
one-forms dual to the $Q_\alpha$ and $P_ \mu$ superalgebra
generators respectively. The form $h=db$ is invariant under
supersymmetry transformations, but $b$ is only quasi-invariant: it
cannot be written in terms of $\Pi^\alpha$, $\Pi^\mu$ since $h$ is
a non-trivial $(2p+2)$-CE cocycle \cite{AT89}. However, there is a
form $\tilde b$ on the specific extended superspace $\widetilde
\Sigma$ that differs from $b$ by a total exterior differential and
can be written in terms of the forms $\Pi^\alpha$, $\Pi^\mu$ and
$\Pi_{\mu_1\dots \mu_{p-k} \alpha_1\dots\alpha_k}$ on $\widetilde
\Sigma$. Since the new coordinates $\varphi_{\mu_1\dots \mu_{p-k}
\alpha_1\dots\alpha_k}$ of $\widetilde \Sigma$ are not present in
$d{\tilde b}=h$, they appear trivially in the action.


\section{Another example of the use of extensions: D-branes,
the M5-brane and worldvolume fields/extended superspace
coordinates democracy} \label{Dbranes}

The action of the 10-dimensional D-branes
\cite{CGMNW,BeTo,APS,HS97} contains a one-form $A(\xi)$, the
Born-Infeld field, that is {\it directly} defined on the
worldvolume parametrized by $\xi=(\tau,\sigma^1,\ldots,\sigma^p)$.
Similarly, that of the 11-dimensional M5-brane
\cite{PST,BLNPST,APPS} contains a worldvolume two-form, which we
shall also denote $A(\xi)$. One can use the extended superspaces
$\widetilde \Sigma$ of Sec.~\ref{Nfermionic} to write these forms
on the worldvolume also as pull-backs (by $\phi^*$) of forms
defined on $\widetilde \Sigma$. Since the forms $A(\xi)$ appear
non-trivially in the actions, the same happens to the new
superspace variables if one writes $A(\xi)=\phi^*(A)$, for some
form $A$ constructed from forms on a suitable $\widetilde \Sigma$
\cite{CAIPB00}. This is an example where the additional
coordinates of $\widetilde \Sigma$ appear non-trivially.

Let us consider the case of the type IIA D$p$--branes, with $p$
even (the case of the type IIB D$p$--branes could be treated
similarly \cite{Sakaguchi,CAIPB00}). In the flat case, with
vanishing dilaton field, their action can be constructed entirely
in terms of the forms of the free differential algebra given by
\begin{eqnarray}
   d\Pi^\alpha &=& 0\ ,\nonumber\\
    d\Pi^\mu &=& \frac{1}{2} (C\Gamma^\mu)_{\alpha\beta}
\Pi^\alpha \wedge \Pi^\beta \ ,\nonumber\\
 d\mathcal{F} &=& \Pi^\mu\wedge
(C\Gamma_\mu\Gamma_{11})_{\alpha\beta} \Pi^\alpha \wedge \Pi^\beta
\ , \label{fdaIIA}
\end{eqnarray}
where the first two equations are the MC equations for the $D=10$,
$N=2$ superPoincar\'e algebra, for which the spinors are of Dirac
type as corresponds to the IIA case, and $\mathcal{F}$ is an
invariant two-form given by
\begin{equation}
       \mathcal{F}=dA-B\ ,\quad \quad dB=-\Pi^\mu \wedge
(C\Gamma_\mu\Gamma_{11})_{\alpha\beta} \Pi^\alpha \wedge \Pi^\beta
\ . \label{FIIA}
\end{equation}
Both the form $\mathcal{F}(\xi)$ that appears in the kinetic and
in the (quasi-invariant) WZ term as $\mathcal{F}_{ij}(\xi)$
($\mathcal{F}(\xi)=\frac{1}{2}\mathcal{F}_{ij}(\xi)d\xi^i \wedge
dx^j$) and $A(\xi)$ are forms directly defined on the worldvolume.
The M5-brane case can be treated similarly by replacing
$\mathcal{F}(\xi)$ by the three-form $H(\xi)=dA(\xi)-C(\xi)$.

If one can find forms $\mathcal{F}$ and $H$ on a suitably extended
superspace $\widetilde \Sigma$ such that their differentials
coincide with those of Eq.~(\ref{fdaIIA}) and with the
corresponding ones for the M5-brane respectively, it follows that
in both cases $A(\xi)$ is $A(\xi)=\phi^*(A)$, where $A$ is
obtained \cite{CAIPB00} by identifying $\phi^*\mathcal{F}$,
$\phi^* H$ with $\mathcal{F}(\xi)$, $H(\xi)$ respectively. The
form $A$ on $\widetilde \Sigma$ contains additional coordinates of
$\widetilde \Sigma$, which are included in $\mathcal{F}$ (or $H$)
inside a total derivative. This is achieved using an extended
superspace $\widetilde \Sigma$, which for the fivebrane is a
$D=11$, $p=2$ extended supergroup (obtained from Eq. \ref{pbralg}
for $p=2$), and for the case of the D--branes is its dimensional
reduction to $D=10$ \cite{CAIPB00}.

The extended superspace $\widetilde \Sigma$ that allows us to
describe the Born-Infeld fields also in terms of one-forms on
$\widetilde \Sigma$  may not be always sufficient (as it is for
the D2--branes) to make the D--brane WZ terms strictly invariant.
It may be seen (see \cite{CAIPB00} for details) that a larger
extended superspace will trivialize the CE $(2p+2)$-cocycles
although it may correspond to a rather large superalgebra (see
\cite{Malg}).

The replacement of $A(\xi)$ by $\phi^*(A)$ in the D--brane and
fivebrane actions gives models that are classically equivalent to
the original ones. This may be seen by noticing that the field
equations obtained by varying the original superspace $\Sigma$
variables and $A(\xi)$ coincide with those obtained by varying the
extended superspace $\widetilde{\Sigma}$ variables in the new
action, provided that the induced worldvolume metric is
non-degenerate, as it is the case in brane theory. Furthermore, it
may be seen \cite{AI01} that there exist the necessary gauge
invariances to reduce the number of degrees of freedom of
$[\phi^*(A)]_i$ (resp.~$[\phi^*(A)]_{ij}$) to those of $A_i(\xi)$
(resp. $A_{ij}(\xi))$.

The above facts support the {\it worldvolume fields/superspace
variables democracy} hypothesis \cite{CAIPB00,AI01}, according to
which {\it the action of the flat superspace version of
superbranes may be written entirely in terms of invariant
one-forms defined on a suitably extended superspace
$\widetilde{\Sigma}$ group.} The fact that D--branes include the
dilaton field in their action does not contradict this conjecture
because the dilaton field in 10 dimensions comes from the K-K
reduction of the 11-dimensional metric, and so it may be viewed as
an effect of moving to a curved $D=11$ spacetime. There is also an
auxiliary (PST) scalar field \cite{PST} in the M5-brane action of
\cite{BLNPST}, its only role in the covariant action being to
account for the required worldvolume self-duality of $A$.

\section{Applications of Lie algebra expansions}
\label{physexp}

\subsection{The complete M-theory superalgebra}
\label{FMthe}

The statement that the M-theory superalgebra is a contraction of
$osp(1|32)$ actually refers to what may be called the `maximal
graded translation algebra' $\Sigma^{(528|32)}$ (in general,
$\Sigma^{(\frac{n(n+1)}{2}|n)}$) . This has a central extension
(of $\{ Q_\alpha,Q_\beta\}=0$ by $[P_{\alpha\beta},
P_{\gamma\delta}]=0$) structure and is given by
\begin{equation}
\hspace{-2cm}  \{ Q_\alpha,Q_\beta\}= P_{\alpha\beta}\ ,\quad
[Q_\alpha, P_{\beta\gamma}]=0 \quad , \quad
P_{\alpha\beta}=P_{\beta\alpha} \ , \quad \alpha, \beta
=1,\ldots,32 \; , \label{mT}
\end{equation}
the generators $P_{\alpha \beta}$ being central. These may be
written as $P_{\alpha\beta}= P_\mu (C\Gamma^\mu)_{\alpha\beta}
+Z_{\mu_1\mu_2} (C\Gamma^{\mu_1\mu_2})_{\alpha\beta}+
Z_{\mu_1\dots \mu_5} (C\Gamma^{\mu_1\dots \mu_5})_{\alpha\beta}$
which is the most general splitting for the symmetric
$P_{\alpha\beta}$ in terms of $Spin(1,10)$ gamma matrices; this
expression breaks the general $GL(32, \mathbb{R})$ invariance of
Eq.~(\ref{mT}) down to $Spin(1,10)$. This M-algebra, however, does
not include the Lorentz automorphisms part.

The MC equations of $osp(1|32)$ may be written as follows:
\begin{eqnarray}
     d\rho^{\alpha\beta} &=& -{\rho^\alpha}_\gamma\wedge
\rho^{\gamma\beta}-\nu^\alpha\wedge \nu^\beta \nonumber\\
d\nu^\alpha &=& - {\rho^\alpha}_\beta\wedge \nu^\beta \ , \quad
\alpha,\beta=1,\ldots,32 \quad, \label{MCosp132}
\end{eqnarray}
where the forms $\rho^{\alpha\beta}=\rho^{\beta\alpha}$ dual to
$Z_{\alpha\beta}$ are bosonic, and those $\nu^\alpha$ dual to
$Q_\alpha$ are fermionic; the indices $\alpha$, $\beta$ are raised
and lowered by means of the $32\times 32$ charge conjugation
matrix $C_{\alpha\beta}$. Let us perform a generalized W-W
contraction relative to the splitting $osp(1|32)=V_0\oplus
V_1\oplus V_2$, where $V_0=0$, $V_1^*$ is generated by the
$\nu$'s, and $V_2^*$ is generated by the $\rho$'s. Then if $\nu$
and $\rho$ are rescaled as $\nu\mapsto \lambda\nu$, $\rho\mapsto
\lambda^2\rho$ and the limit $\lambda \rightarrow 0$ is taken, one
arrives at
\begin{eqnarray}
     d\rho^{\alpha\beta} &=& -\nu^\alpha\wedge \nu^\beta
\nonumber\\ d\nu^\alpha &=& 0\ ,  \label{MCosp132c}
\end{eqnarray}
which is precisely the dual or MC forms version of the
superalgebra (\ref{mT}). The dimensions of both $osp(1|32)$ and
the maximal graded translations algebra in $D=11$, Eq.~(\ref{mT}),
are the same: $32\times 33/2+32=560$. However, the full M-theory
superalgebra has the additional ${11 \choose 2}=55$ Lorentz
generators, so it is not possible to obtain it by contracting the
smaller $osp(1|32)$ algebra.

Nevertheless, the M-theory superalgebra including the Lorentz part
can be obtained as the expansion $osp(1|32)(2,1,2)$ of
$osp(1|32)$. Let us start by splitting $osp(1|32)=V_0\oplus
V_1\oplus V_2$, where now the (dual) space $V_0^*$ is generated by
the $\rho^{\mu\nu}$ and $V_2^*$ is generated by $\rho^\mu$ and
$\rho^{\mu_1\dots \mu_5}$. This is made explicit by writing
\begin{equation}
\hspace{-2.4cm} \rho_{\alpha\beta}= -\frac{1}{32} \left( \rho_\mu
C\Gamma^\mu -\frac{1}{2} \rho_{\mu\nu} C\Gamma^{\mu\nu}+
\frac{1}{5!} \rho_{\mu_1\dots \mu_5} C\Gamma^{\mu_1\dots
\mu_5}\right)_{\alpha\beta} \; , \;\; \mu,\nu=0,1,\ldots,10 \;
.\label{generalrho}
\end{equation}
If, fulfilling condition (\ref{consup2}), we set $N_0=2$, $N_1=1$,
$N_2=2$, this means that we expand $\rho$ and $\nu$ as follows:
\begin{equation}
\hspace{-1.8cm} \nu=\lambda\nu ^{(1)}\ ,\quad
\rho_{ab}=\rho_{\mu\nu}^{(0)}+\lambda^2 \rho_{\mu\nu}^{(2)}\
,\quad \rho_\mu =\lambda^2 \rho_\mu^{(2)} \ ,\quad
\rho_{\mu_1\dots \mu_5}=\lambda^2 \rho_{\mu_1\dots \mu_5}^{(2)}\ .
\label{fullMexp}
\end{equation}
It is then seen that the new MC equations are precisely the dual
of the complete M-theory superalgebra, the Lorentz generators
being $\rho_{\mu\nu}^{(0)}$, and the `generalized translations'
being $\rho_\mu^{(2)}$, $\rho_{\mu\nu}^{(2)}$, $\rho_{\mu_1\dots
\mu_5}^{(2)}$, which can be collected as
$\rho_{\alpha\beta}^{(2)}$. Therefore, using the notation of the
introduction, it follows that the full M-theory superalgebra is
$osp(1|32)(2,1,2)$.

\subsection{Expansions of gauge differential algebras and Chern-Simons
Poincar\'e supergravity in $2+1$ dimensions} \label{poinsup}

It is known that Poincar\'e supergravity in $2+1$ dimensions is a
Chern-Simons (CS) gauge theory based on the superPoincar\'e
algebra. It can also be shown that it may be obtained from a
(contraction) limit (that setting the cosmological constant equal
to zero) of the $(2+1)$-dimensional type $(0,1)$ anti-de Sitter
supergravity \cite{AchT}, which is also a CS theory based on
$sp(2)\oplus osp(1|2)$ (the IW contraction limit involves
simultaneously the two algebras). We are going to show here that
Poincar\'e supergravity in $D=3$ may also be obtained from an
expansion of a CS model based on $osp(1|2)$, with an appropriate
splitting. This is based on the fact that the expansion method may
be also used to expand the gauge theories associated with the
original algebra \cite{AIPV}.

Let us start from the MC equations of $osp(1|2)$. These are given
also by (\ref{MCosp132}), but now with $\alpha,\beta=1,2$. The
corresponding gauge free differential algebra (FDA) is given in
terms of the gauge potentials $f^{\alpha\beta}$ and $\xi^\alpha$
and their curvatures $\Omega^{\alpha\beta} =df^{\alpha\beta}+
{f^\alpha}_\gamma \wedge f^{\gamma\beta}+\xi^\alpha\wedge
\xi^\beta$ and $\Psi^\alpha=d\xi^\alpha +{f^\alpha}_\beta \wedge
\xi^\beta$ by the equations defining the curvatures and the
Bianchi identities. Using this FDA, one sees that the gauge
invariant $4$-form
\begin{equation}
{\cal H} ={\Omega^\alpha}_\beta \wedge{\Omega^\beta}_\alpha -2
\Psi_\alpha \wedge\Psi^\alpha \label{HCS3}
\end{equation}
is closed. So, if ${\cal B}$ is its CS form, $d{\cal B}={\cal H}$,
it is possible to define a CS model through the action
$\int_{\mathcal{M}^3} {\cal B}$.

Let us split $osp(1|2)$ in the form $osp(1|2)=V_0\oplus V_1$,
where the dual space $V_0^*$ contains the one-forms
$\rho^{\alpha\beta}$, and $V_1^*$ contains the $\nu^\alpha$. It
may be shown \cite{AIPV} that the expansion of the gauge
potentials follows the same pattern as that of the MC forms,
\begin{equation}
     f^{\alpha\beta} = \sum^\infty_{n=0}
f^{\alpha\beta,2n}\lambda^{2n} \ ,\quad \xi^\alpha =
\sum^\infty_{n=0} \xi^{\alpha,2n+1}\lambda^{2n+1}\
\label{gaugexp3}
\end{equation}
and similarly for $\Omega^{\alpha \beta}$ and $\Psi^\alpha$. We
now assign physical dimensions to the parameter $\lambda$. Since
we want to make contact with gravity, we would like
$f^{\alpha\beta,0}$ to correspond to the Lorentz generators and
$f^{\alpha\beta,2}$ to the dreibein forms. This means that
$[\lambda]=L^{-1/2}$. On the other hand, the action for $D=3$
gravity in geometrized units has dimensions of $L$, so if we
expand the CS action in $\lambda$ we need the term in $\lambda^2$
in it in order to obtain a new CS action with the right physical
dimensions. The resulting action and its corresponding
superalgebra $osp(1|2)(2,1)$ (the consistent one that contains all
the gauge fields that appear in the action integrand), coincides
with the $D=3$ supergravity action and the $D=3$ superPoincar\'e
algebra respectively. Indeed, the $osp(1|2)$ action is
\begin{equation}
\hspace{-2.5cm} \int_{\mathcal{M}^3} {\cal B}= \int_{{\cal
M}^3}\left(f^\alpha{}_\beta \wedge \Omega^{\beta}{}_\alpha
-2\xi_\alpha \wedge \Psi^\alpha -\frac13 f^\alpha{}_\beta \wedge
f^\beta{}_\gamma \wedge f^\gamma{}_\alpha - f^\alpha{}_\beta
\wedge \xi^\beta \wedge \xi_\alpha \right) \; .
     \label{osp12CS}
\end{equation}
Inserting the expansions (\ref{gaugexp3}), selecting the
$\lambda^2$ terms and using that, in three dimensions, one may
write
\begin{eqnarray}
& f^{\alpha\beta,0}=
\frac{1}{4}(C\Gamma^{ab})^{\alpha\beta}\omega_{ab}\ ,\
\Omega^{\alpha\beta,0}=
\frac{1}{4}(C\Gamma^{ab})^{\alpha\beta}R_{ab}\ , \nonumber \\
& f^{\alpha\beta,2}= -\frac{1}{2}(C\Gamma^a)^{\alpha\beta}e_a\ , \
\xi^{\alpha,1} = \psi^\alpha \ , \label{osppoin}
\end{eqnarray}
we obtain the $D=3$ superPoincar\'e gravity action,
\begin{equation}
       I= \int_{{\cal M}^3}\left(\epsilon^{abc}R_{ab}\wedge
e_c+ 4\psi_\alpha \wedge D\psi^\alpha \right) \; .
     \label{poincareCS}
\end{equation}

The method may be applied to Chern-Simons supergravities in higher
dimensions (see {\it e.g.}~\cite{Zanelli, HN} for an outlook of CS
supergravities and further references) to compare {\it e.g.}, with
the standard supergravity \cite{CJS} and the approach of
\cite{DFR}.

\vspace{0.5cm}
 \noindent {\bf Acknowledgments.} This work has been
partially supported by the Spanish Ministry of Science and
Technology through grants BFM2002-03681, BFM2002-02000 and EU
FEDER funds, and by the Junta de Castilla y Le\'on through grant
VA085-02. Two of the  authors also wish to thank the Spanish
Ministry of Education and Culture (M.P.) and the Generalitat
Valenciana (O.V.) for their research grants.

\end{document}